\documentclass[9pt,twocolumn,twoside]{osajnl}
\journal{ol} 

\setboolean{shortarticle}{true}

\title{Omnidirectional field enhancements drive giant nonlinearities in epsilon-near-zero waveguides}

\author[1]{Gordon Han Ying Li}
\author[1,2]{C. Martijn de Sterke}
\author[1,2]{Alessandro Tuniz}

\affil[1]{Institute of Photonics and Optical Science (IPOS), School of Physics, The University of Sydney, NSW 2006, Australia}
\affil[2]{The University of Sydney Nano Institute (Sydney Nano), The University of Sydney, NSW 2006, Australia}

\affil[*]{Corresponding author: alessandro.tuniz@sydney.edu.au}

\begin{abstract}
Bulk materials possessing a relative electric permittivity $\varepsilon$ close to zero exhibit giant Kerr nonlinearities. However, harnessing this response in guided-wave geometries is not straightforward, due to the extreme and counter-intuitive properties of epsilon-near-zero materials. Here we investigate, through rigorous calculations of the Kerr nonlinear coefficient, how the remarkable nonlinear properties of such materials can be exploited in several different types of structures, including bulk films, plasmonic nanowires, and metal nanoapertures. We find the largest Kerr nonlinear response when both the modal area and the group velocity are simultaneously minimized, corresponding to omnidirectional field enhancement. The physical insights developed will be key for understanding and engineering nonlinear nanophotonic systems with extreme nonlinearities and point to new design paradigms.
\end{abstract}

\setboolean{displaycopyright}{true}

\begin{document}

\maketitle
A long-standing goal of nonlinear optics is the identification and engineering of materials and structures with large nonlinear responses~\cite{byer1974nonlinear}. Large Kerr optical nonlinearities  have numerous applications in quantum information~\cite{chang2014quantum}, sensing~\cite{mesch2016nonlinear}, optical signal processing~\cite{cotter1999nonlinear}, and nonlinear spectroscopy~\cite{mukamel1999principles}; it is advantageous to reduce the footprint of such devices in terms of power consumption and physical dimensions, especially for nanophotonic integration~\cite{niu2018epsilon}. For bulk materials, the Kerr nonlinearity is quantified using the nonlinear refractive index $n_{2}$~\cite{boyd2019nonlinear}
\begin{equation}
n_{2}=\dfrac{3\chi^{(3)}}{4\sqrt{\varepsilon_{m}}\Re {e}(\sqrt{\varepsilon_{m}})\varepsilon_{0}c_0}\ ,
\label{eq:n2}
\end{equation}
where $\varepsilon_{0}$ is the vacuum permittivity, $c_0$ is the speed of light in vacuum, $\chi^{(3)}$ is the third-order nonlinear optical susceptibility, and $\varepsilon_{m}$ is the linear permittivity. In this case, changes in the refractive index vary linearly with the intensity $I$ as $\Delta n=n_{2}I$. Only recently have the nonlinear responses of doped semiconductors~\cite{caspani2016enhanced,alam2016large} and metamaterials~\cite{moitra2013realization,kaipurath2016optically} been characterized in regions where the electric permittivity vanishes (typically at near-infrared frequencies). These show an extremely large $n_{2}$ as $\Re {e}(\varepsilon_{m}) \rightarrow 0$~\cite{caspani2016enhanced}. These so-called $\varepsilon$-near-zero (ENZ) materials have been demonstrated to exhibit a wealth of enhanced nonlinear effects including efficient harmonic generation~\cite{capretti2015comparative} and unprecedentedly large changes in refractive index~\cite{reshef2019nonlinear}. 

ENZ materials such as indium tin oxide (ITO) possess an ultrafast Kerr nonlinearity and are compatible with CMOS fabrication processes~\cite{alam2016large}, making them attractive candidates for novel photonic integrated circuits~\cite{niu2018epsilon}. Indeed, there have been two different methods proposed to extend ENZ behaviour to guided-mode structures: (i) to operate a waveguide with ENZ material inclusions at the bulk ENZ condition $\Re {e}(\varepsilon_{m})=0$~\cite{campione2015theory,vassant2012berreman,campione2016experimental,minn2018excitation}; (ii) to operate a plasmonic waveguide with effective mode permittivity $\varepsilon_{\rm eff}$ near cutoff such that $\Re {e}(\varepsilon_{\rm eff})=0$~\cite{argyropoulos2012boosting,li2019exceptional,li2018epsilon}. 

The Kerr nonlinearity in waveguides is quantified by the nonlinear coefficient $\gamma$, whereby changes in the propagation constant $\kappa$ vary linearly with the power $P$ as $\Delta\kappa=\gamma P$. The expression for  $\gamma$ for extremely lossy waveguides is given by~\cite{li2017general,li2020establishing}:

\begin{equation}
\gamma =\dfrac{3\omega\varepsilon_{0}}{4\Re e\left[\int_{-\infty}^{\infty}(\mathbf{e}\times\mathbf{h}^{*})\cdot\hat{z}\ dxdy\right]}\dfrac{\int_{-\infty}^{\infty}\chi^{(3)}|\mathbf{e}|^{2}\left(\mathbf{e}\cdot\mathbf{e}-2e_{z}^{2}\right)\ dxdy}{\int_{-\infty}^{\infty}(\mathbf{e}\times\mathbf{h})\cdot\hat{z}\ dxdy},
\label{eq:gamma_Li}
\end{equation}
where $\mathbf{e}$, $\mathbf{h}$ are electric- and magnetic- modal fields respectively, $\omega$ is the angular frequency, $\hat z$ points in the longitudinal (propagation) direction, and the $xy$ plane is transverse. For 1D waveguides, Eq.~\ref{eq:gamma_Li} is integrated over $\hat{x}$ (transverse).

In this work we evaluate $\gamma$ for several ENZ geometries and elucidate the physics responsible for enhanced Kerr nonlinearities in waveguides. We analyze the giant nonlinearity in bulk ENZ media in terms of the group velocity, and analyze the nonlinear coefficient of several representative structures at frequencies close to the ENZ frequency: a semi-infinite metal/dielectric film, a metal nanowire, and a nano-aperture. We find that the largest nonlinear response occurs when \textit{both} transverse- and longitudinal- field enhancements occur, corresponding to low group velocities and small modal areas, respectively, obtained simultaneously using structured metallic media. This general framework  
opens the door to novel approaches in compact, energy-efficient nanostructures with giant nonlinear responses.

\begin{figure}[t!]
\centering
\includegraphics[width=0.45\textwidth]{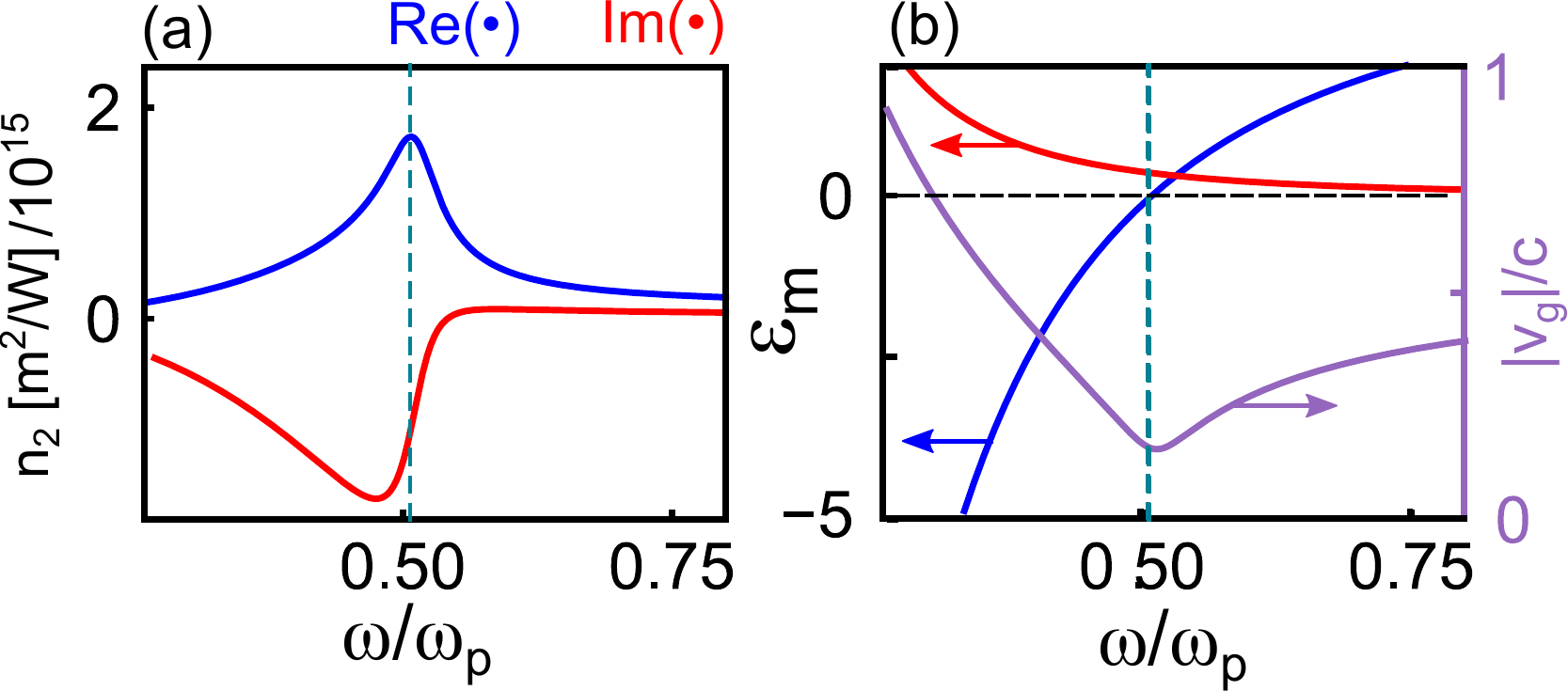}
\caption{(a) Real- (blue) and imaginary- (red) parts of $n_2$ for bulk ITO in air. (b) Associated real- (blue) and imaginary- (red) parts of $\varepsilon_m$ (left axis), and $|v_g|$ of bulk ITO (right axis). The condition $\Re e (\varepsilon_m) \sim 0$ (horizontal dashed line) and small $|v_g|$ both correspond to a maximum $\Re e(n_2)$ (vertical dashed line).}
\label{fig:fig1}
\end{figure}

While our discussion is valid for any metal near its ENZ frequency, we consider ITO with linear Drude relative permittivity
\begin{equation}
\varepsilon_{m}(\omega)=\varepsilon_{\infty}-\dfrac{\omega_{p}^{2}}{\omega^{2}+i\Gamma\omega}\ ,
\label{eq:drude}
\end{equation}
where $\varepsilon_{\infty}$ is its high-frequency limit, $\omega_{p}$ is the plasma frequency, and $\Gamma$ is the damping rate. For bulk materials, the ENZ frequency for which $\Re {e}(\varepsilon_{m})=0$ is $\omega=(\omega_{p}^{2}/\varepsilon_{\infty}-\Gamma^{2})^{1/2}$.  We take ITO to be the nonlinear medium ($\varepsilon_{\infty}=3.8055$, $\omega_{p}=2\pi\times 473~\mathrm{THz}$, and $\Gamma=2\pi\times 22~\mathrm{THz}$~\cite{alam2016large}), possessing a third-order nonlinear optical susceptibility $\chi^{(3)}=(1.6+0.5i)\times 10^{-18}~\mathrm{m^{2}/V^{2}}$~\cite{alam2016large}. For simplicitly we take $\chi^{(3)}$ to be constant, since the giant nonlinear response of ENZ materials is not due to changes in $\chi^{(3)}$~\cite{caspani2016enhanced}. We assume air to be the surrounding linear dielectric with constant permittivity $\varepsilon_{d}=1$. 
The complex $n_{2}$ of ITO, calculated using Eq.~\ref{eq:n2}, is shown in Fig.~\ref{fig:fig1}(a), and the relative permittivity of ITO is shown in Fig.~\ref{fig:fig1}(b).  The ENZ frequency (vertical dashed line), is indeed optimal for enhancing $\Re e({n_2})$ of bulk ITO, consistent with previous studies~\cite{alam2016large}. It was recently suggested that the physical origin of this enhanced nonlinearity is a local minimum in the group velocity  $v_g$, because $v_g/\sqrt{\varepsilon_m} = c_0/\varepsilon_\infty$ if $\Gamma \ll \omega$~\cite{kinsey2019nonlinear}. Thus, the enhancement in $n_2$ in the ENZ region could be interpreted as originating from a small $v_g$~\cite{kinsey2019nonlinear} that enhances the electric field. Figure~\ref{fig:fig1}(b) shows the calculated $|v_g| = |d\omega/d\kappa|$ for bulk ITO, confirming that the minimum $v_g$ also coincides with the largest $n_2$. The Kerr nonlinear effect can be further improved for angled incidence~\cite{alam2016large} and transverse magnetic (TM) polarization: the continuity of the electric displacement perpendicular to the film leads enhances the associated field by a factor of $\varepsilon_d/\varepsilon_m$ inside the ENZ medium, and thus to enhanced nonlinearities. These omnidirectional field enhancements in unison could therefore drive even larger Kerr nonlinearities, particularly in ENZ waveguides.

\begin{figure}[b!]
\centering
\includegraphics[width=0.45\textwidth]{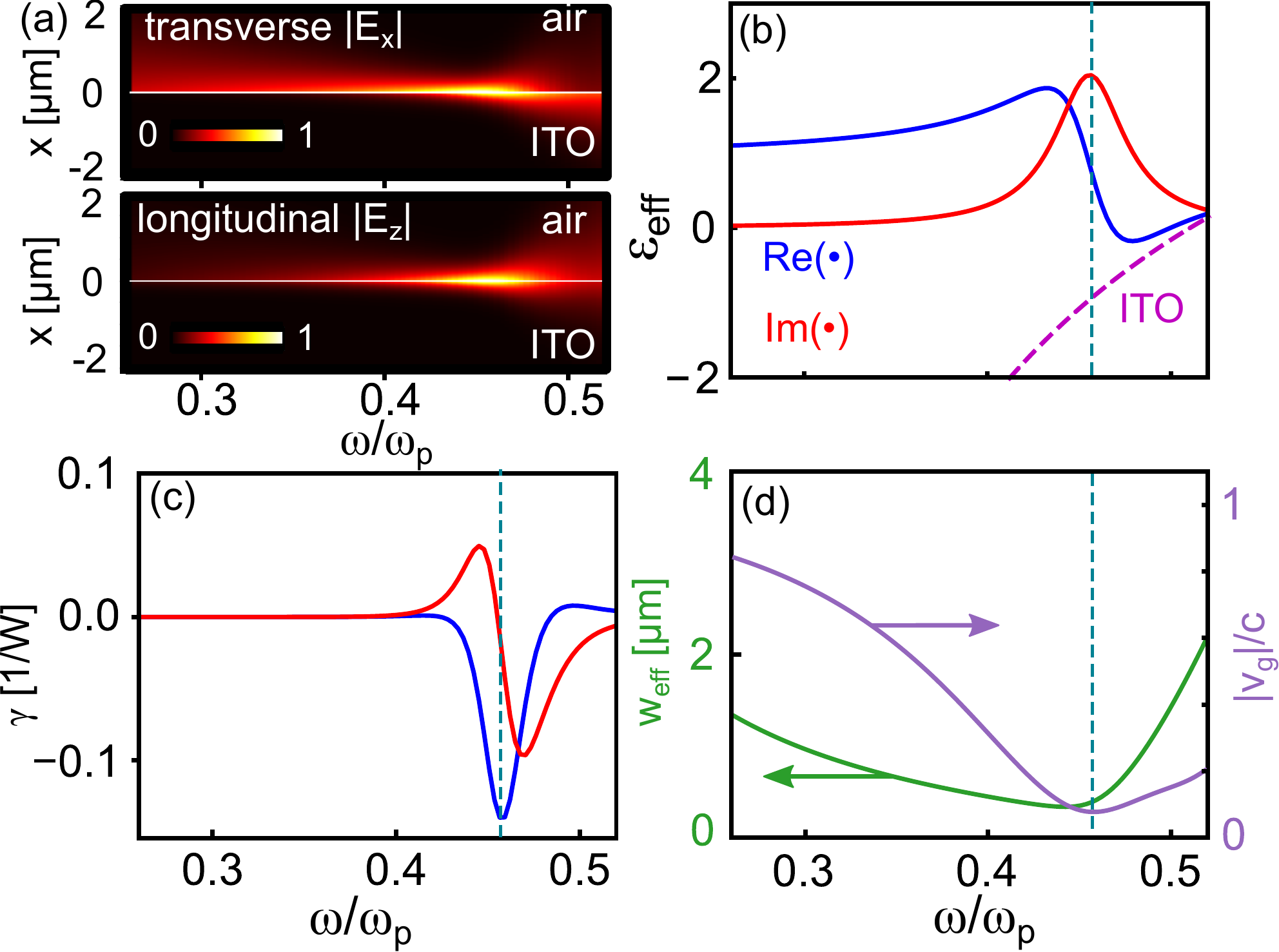}
\caption{(a) Calculated longitudinal and transverse $|E|$ for an ITO/air SPP propagating in $\hat{z}$ as a function of $\omega/\omega_p$. $\hat{x}$ is transverse. All fields carry the same power. (b) Real- and imaginary- parts of $\varepsilon_{\rm eff}$ of the SPP mode as defined by Eq.~\ref{eq:kappa} (blue- and red- lines, respectively). $\Re e(\varepsilon_m)$ is shown as a dashed line for comparison. (c) Real- and imaginary- parts of $\gamma$. (d) Associated $w_{\rm eff}$ ($1/e$ width of $|E|$) (green) and $|v_g|/c_0$ (purple). Vertical dashed line: frequency at maximum $|\Re e (\gamma)|$.}
\label{fig:fig2}
\end{figure}

With this insight, we now consider the nonlinear response of structured waveguides formed by ITO, with an eye on using such structures as highly nonlinear modules within compact nonlinear photonic integrated circuits~\cite{tuniz2020modular}. We start by analyzing the simple case of a surface plasmon polariton (SPP) mode of a semi-infinite ITO/air interface, for which closed expressions for the propagation constants and field components are known~\cite{maier2007plasmonics}. The frequency-dependent longitudinal and transverse field magnitudes of such modes are shown in Fig.~\ref{fig:fig2}(a). The dispersion relation of SPP modes is given by
\begin{equation}
\kappa=k_{0} \sqrt{\varepsilon_{\rm eff}} = k_{0}\sqrt{\dfrac{\varepsilon_{m}\varepsilon_{d}}{\varepsilon_{m}+\varepsilon_{d}}}\ ,
\label{eq:kappa}
\end{equation}
where $k_{0}$ is the vacuum wave number. Figure~\ref{fig:fig2}(b) shows the real- and imaginary- parts of $\varepsilon_{\rm eff}$ as a function of $\omega/\omega_p$. The relative permittivity of bulk ITO is shown as a dashed line for comparison. 
The closed form expression for $\gamma$ in SPPs~\cite{li2020establishing} has a more complicated form than the bulk case (Eq.~\ref{eq:n2}). The resulting $\gamma$ is plotted in Fig.~\ref{fig:fig2}(c): we observe that the largest nonlinearity does not occur at the material's ENZ condition, but rather where the longitudinal and transverse fields are both large, as can be seen in Fig.~\ref{fig:fig2}(a). To better address the underlying physical mechanisms, we now revisit the nonlinear coefficient $\gamma$.

 Equation~\ref{eq:gamma_Li} represents $\gamma$  for arbitrary lossy waveguides, however extracting physical meaning from it is challenging. Afshar et al.~\cite{afshar2013understanding} proposed a factorization of $\gamma$ for lossless waveguides in terms of physically meaningful quantities such as $v_g$ and effective modal area $A_{\rm eff}$. Following that approach, we suggest the following heuristic relationship for lossy waveguides:

\begin{equation}
\gamma\propto\dfrac{k_{0}\cdot\chi^{(3)}}{v_{g}^2\cdot A_{\rm eff}}\ .
\label{eq:gamma}
\end{equation}
In the 1D case, $A_{\rm eff}$ in Eq.~\ref{eq:gamma} is replaced by an effective width $w_{\rm eff}$. As we will show, Eq.~\ref{eq:gamma} proves useful for understanding the origin of enhanced Kerr nonlinearities in waveguides formed by ENZ media. Briefly, the rationale behind Eq.~\ref{eq:gamma} is as follows. Firstly, $\gamma\propto\chi^{(3)}$~\cite{li2020establishing}, 
and $\gamma\propto k_{0}$ since $k_{0}$ relates the phase and propagation distance. We expect that $\gamma\propto 1/v_{g}^2$ since a low group velocity enhances the electric field \cite{Monat2010review}.This slow light is a purely longitudinal effect that causes the trailing edge of a pulse's field to catch up with its leading edge as the group velocity decreases~\cite{Monat2010review, miyata2012field}. Similarly, we expect that $\gamma\propto 1/A_{\rm eff}$ since a small effective modal width enhances the electric field via transverse confinement.
In other words, omnidirectional enhancement of the electric field would lead to giant Kerr nonlinearities, whereby the maximum $\gamma$ occurs when $v_{g}$ and $A_{\rm eff}$ are minimized simultaneously.

While the simultaneous minimization of $v_{g}$ and $A_{\rm eff}$ is not necessarily achieved in all-dielectric waveguides, even the simple 1D SPP mode satisfies both conditions within a narrow frequency region, close to the point of the (lossless) electrostatic surface plasmon polariton (ESPP) ($\varepsilon_{m}=-\varepsilon_{d}$)~\cite{maier2007plasmonics}: such ESPP modes exhibit a divergent propagation constant at a finite frequency, thereby vanishing group velocity and extreme field enhancement. Additionally, the transverse wavevectors $k_t$ in both media take on a maximum value since $k_t^2+\kappa^2=k_0^2\epsilon_{m,d}$, leading to strong transverse spatial confinement. The introduction of losses leads to a resonant effective permittivity dispersion near $\Re e(\varepsilon_{m})\approx-\varepsilon_{d}$ that produces a local maximum in the propagation constant and a local minimum in the group velocity. Figure~\ref{fig:fig2}(d) shows the magnitude of $v_{g}$ and $w_{\rm eff}$ as a function of frequency for the 1D SPP: $|v_{g}|$ decreases monotonically and reaches a minimum around the ESPP frequency, before increasing above it. The $w_{\rm eff}$ displays a similar trend, and the maximum $|\Re e(\gamma)|$ is close to these minima. In more complicated geometries the argument is similar: in the electrostatic regime the longitudinal field is compressed around the metal-dielectric interface to the degree that the remaining waveguide elements do not matter. 

\begin{figure}[t!]
\centering
\includegraphics[width=0.45\textwidth]{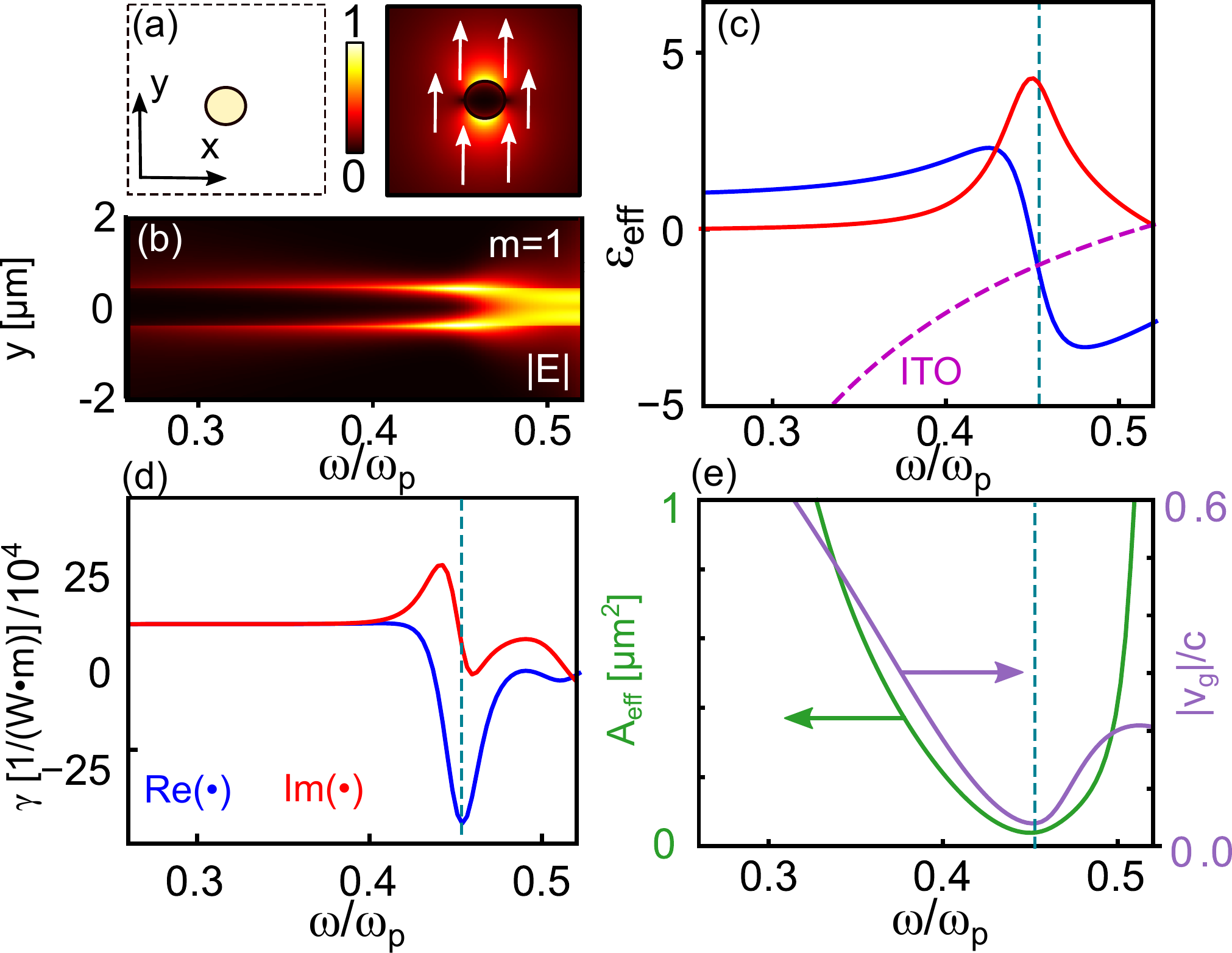}
\caption{(a) Schematic of a the linearly polarized mode of an ITO nanowire (radius: 400\,nm). (b) Calculated $|E|$ at the center vertical axis as a function of $\omega/\omega_p$. (c) Real- and imaginary- parts of $\varepsilon_{\rm eff}$. Dashed line: $\Re e (\varepsilon_m)$. (d) Calculated real- and imaginary- parts of $\gamma$ (Eq.~\ref{eq:gamma_Li}). (e) Associated $A_{\rm eff}$ (green) and $|v_g|/c_0$ (purple). Vertical dashed line: maximum $|\Re e (\gamma)|$.}
\label{fig:fig3}
\end{figure}
The ESPP mode is not unique to semi-infinite metal-dielectric interfaces, and a similar overall behaviour is expected for other nanostructured plasmonic waveguides. We now consider two representative 2D nanoscale waveguides structured with ITO, formed by cylindrical plasmonic structures whose Eigenmodes are obtained from numerical solutions of an analytical transcendental equation~\cite{schmidt2008long} for rapid and accurate parameter sweeps. We first analyze a cylindrical ITO nanowire of radius 400\,nm surrounded by air, and solve for unity azimuthal order ($m=1$), resulting in the frequency-dependent linearly polarized cylindrical modes shown in Fig.~\ref{fig:fig3}(a). These modes can be interpreted as the cylindrical counterpart of the bulk SPP as shown in Fig.~\ref{fig:fig3}(b), which shows the frequency-dependent electric field norm along $y$, centered in $x$. 
The resulting effective waveguide permittivity $\varepsilon_{\rm eff}$ and $\gamma$ (Eq.~\ref{eq:gamma_Li}) are shown in Fig.~\ref{fig:fig3}(b) and ~\ref{fig:fig3}(c) respectively, exhibiting similar features to the SPP: a resonant effective permittivity around the ESPP frequency, where $|\Re e(\gamma)|$ is maximum. The resulting $v_g$ and $A_{\rm eff}$ are shown in Fig.~\ref{fig:fig3}(c): again, both values have a global minimum at the point where the nonlinear coefficient is maximum. Here we have chosen $A_{\rm eff}$ as the area of the power flow of the plasmonic Eigenmode in the propagation direction (Eq. (34) in Ref.~\cite{afshar2013understanding}). While there is no unique definition of $A_{\rm eff}$, we find the results presented here to be qualitatively independent of this choice, because all definitions of $A_{\rm eff}$ inform on the degree of transverse confinement. The enhancement of $\gamma$ at the ESPP frequency is far greater compared to that at the ENZ frequency due to the more drastic reductions in $v_{g}$ and $A_{\rm eff}$, analogously to the SPP mode of the ITO-air interface.

\begin{figure}[t!]
\centering
\includegraphics[width=0.45\textwidth]{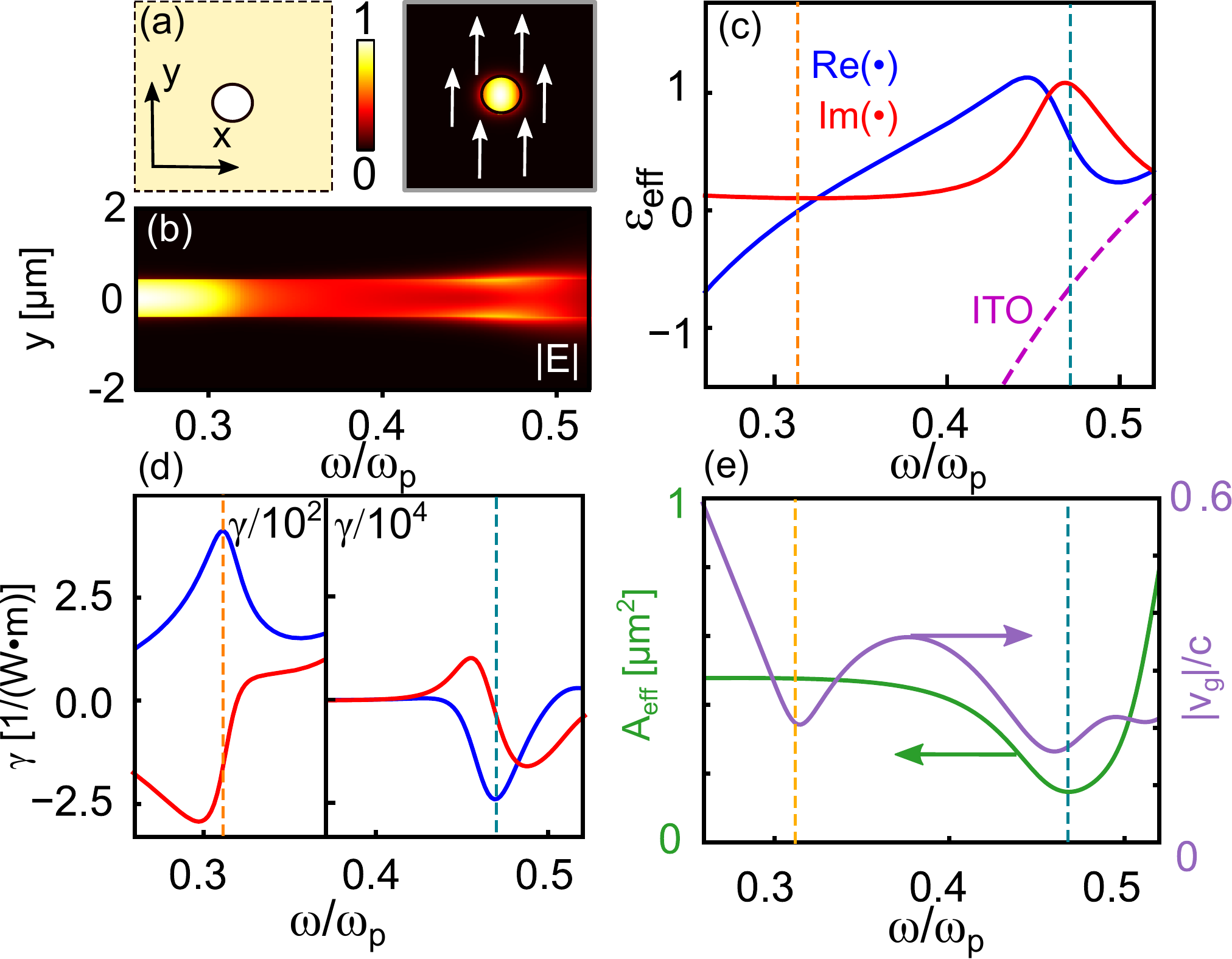}
\caption{(a) Schematic of a the linearly polarized mode of an air/ITO nanoaperture (radius: 400\,nm). (b) Calculated $|E|$ at the center vertical axis as a function of $\omega/\omega_p$. (b) Real- and imaginary- parts of $\varepsilon_{\rm eff}$. Dashed line: $\Re e (\varepsilon_m)$. (c) Calculated real- and imaginary- parts of $\gamma$ (Eq.~\ref{eq:gamma_Li}). (d) Associated $A_{\rm eff}$ (green) and $|v_g|/c_0$ (purple). Vertical dashed lines correspond to local maxima of $|\Re e (\gamma)|$.}
\label{fig:fig4}
\end{figure}

As a final example, we calculate the properties of the inverse geometry, formed by air/ITO nano-apertures of radius 400\,nm, illustrated in the Fig.~\ref{fig:fig4}(a) schematic. Here the field is mostly confined inside the aperture over the entire wavelength region as shown in Fig.~\ref{fig:fig4}(b). In addition to the resonant behaviour near the ESPP frequency, the effective permittivity (shown in Fig.~\ref{fig:fig4}(b)) also possesses $\Re e(\varepsilon_{\rm eff}) =0 $ at $\omega/\omega_p = 0.31$ (dashed yellow line), corresponding to the cutoff frequency of the metal aperture in the complex propagation plane~\cite{novotny1994light}. Here $|\Re e(\gamma)|$ (shown in Fig.~\ref{fig:fig4}(c)) has a local maximum at this aperture cutoff frequency and a global maximum near the ESPP frequency. The associated $v_g$ and $A_{\rm eff}$ are shown in Fig.~\ref{fig:fig4}(d): the local maximum $\gamma$ at $\omega/\omega_p = 0.31$ corresponds to a local minimum $v_g$. This aperture cutoff uniquely lowers $v_g$, with $A_{\rm eff}$ staying nearly constant in the aperture at low frequencies. The global maximum of $|\Re e(\gamma)|$ is again near the global $v_g$ and $A_{\rm eff}$ minima.

Given our improved physical picture, we can revisit the example of bulk ENZ media. We had earlier associated the small group velocity $v_{g}=\sqrt{\varepsilon_{m}}c/\varepsilon_{\infty}$ at the ENZ frequency with a large transverse field enhancement. An analysis of the waveguide case shows that this can be further enhanced with longitudinal field enhancement, which in the case of a bulk film was addressed with angled beams~\cite{alam2016large}. The enhanced Kerr nonlinearity in both bulk ENZ media and guided-wave structures can both be understood in the same framework of a omnidirectional field enhancement. This insight clarifies the underlying physics and shifts the focus away from pursuing a mathematical zero-permittivity condition.

\begin{figure}[t!]
\centering
\includegraphics[width=0.45\textwidth]{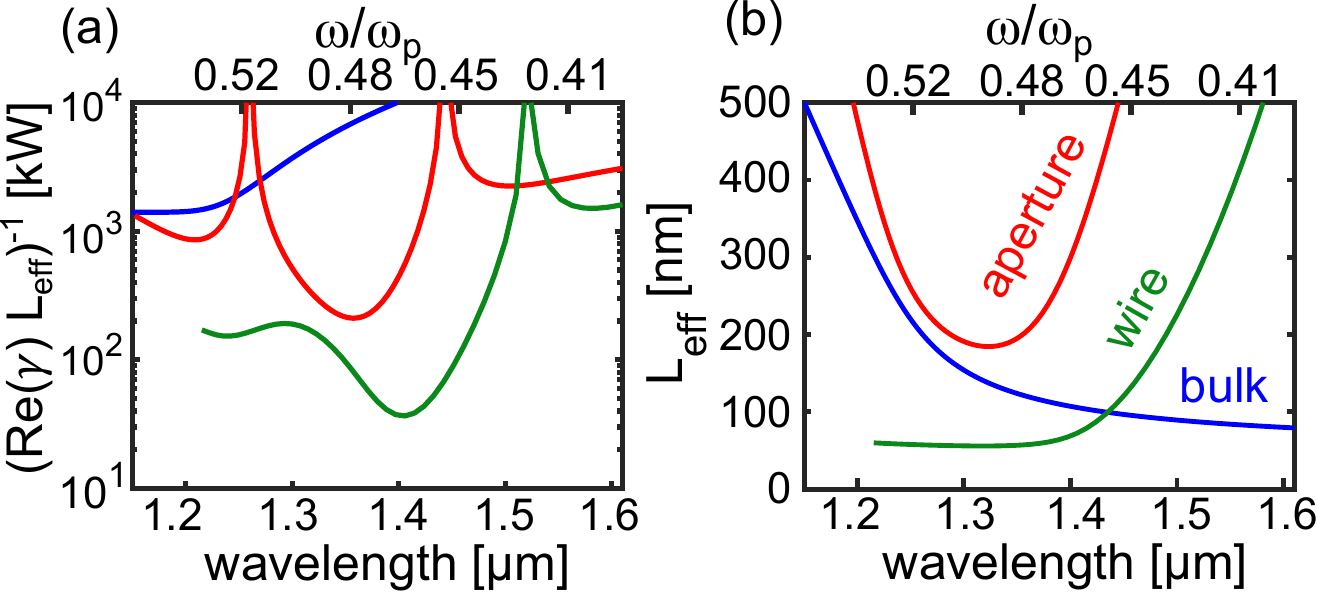}
\caption{(a) (FOM)$^{-1}$ and (b) $L_{\rm eff}$ for ITO (blue), the nanowire in Fig.~\ref{fig:fig3} (green), and the nanoaperture in Fig.~\ref{fig:fig4} (red).}
\label{fig:fig5}
\end{figure}

We now comment on the nonlinear performance of these structured nonlinear devices compared to bulk ITO, which itself already exhibits giant Kerr nonlinearities~\cite{alam2016large}. For comparison, we consider the upper limit of the nonlinear coefficient of bulk ITO to be $\gamma_{\rm ITO} = k_0 n_2/A_{\rm eff,0}$, where $A_{\rm eff, 0} = 2\times\pi w_0^2$~\cite{tuniz2020pulse} is obtained by applying the definition of $A_{\rm eff}$ to a Gaussian beam with $1/e$ radius of $w_0 = 0.61\lambda$, corresponding to the diffraction limit. A commonly figure of merit (FOM) is the normalized nonlinear phase shift $\Re {e}(\gamma) L_{\rm eff}$~\cite{boyd1989applications}, where $L_{\rm {eff}} = [2 k_0 \Im m (\sqrt{\varepsilon_{\rm eff}})]^{-1}$ is the effective propagation length in the waveguide, so that $[\Re {e}(\gamma)L_{\rm eff}]^{-1}$ is the nominal input power required to achieve a nonlinear phase shift of one radian. 
Figure~\ref{fig:fig5}(a) compares $[|\Re {e}(\gamma)|L_{\rm eff}]^{-1}$ for a bulk ITO film with that of the nanowires and nano-apertures calculated in Fig.~\ref{fig:fig3} and Fig.~\ref{fig:fig4}, respectively, and Fig.~\ref{fig:fig5}(b) shows the effective propagation length scales associated with this FOM. Considering bulk ITO as reference, a one radian phase shift at maximum $n_2$ (i.e., at $\lambda = 1.24\,\mu{\rm m}$) is achieved with 1.7\,KW over an effective length $L_{\rm eff} = 240\,{\rm nm}$, corresponding to $\sim 50\,{\rm GW/cm}^2$, in agreement with typical experimental configurations~\cite{alam2016large}. For the nanowire considered, we find a 46-fold reduction in power at $\lambda = 1.4\,\mu{\rm m}$; for the nanoaperture, we find an 8-fold power reduction at $\lambda = 1.36\,\mu{\rm m}$. While the nanoaperture possesses a comparable $L_{\rm eff}$ to bulk, the nanowire's effective length is reduced by a further factor of three. From an experimental perspective, these two characteristics indicate that appropriately engineered structured ITO metasurface arrays (e.g., pillars and nanoholes) boost the already giant Kerr nonlinear responses of ENZ media over much smaller length scales, leading to an order-of-magnitude improvement in performance when accounting for both physical footprint and power reduction combined. Note that a full description of the nonlinear phase shift requires a full non-perturbative treatment~\cite{reshef2017beyond}, with the performance of the device also being linked to the coupling efficiency. Although the starting point of our treatment was the nonlinear coefficient of a propagating waveguide mode, in ultimate analysis the waveguides should have deep sub-wavelength thickness ($\sim \lambda/ 30$), which naturally links to the nonlinear metasurfaces formed by ENZ media. Preliminary experiments on similar structures have been promising~\cite{guo2016large}, and the origin of their nonlinearity was analyzed using a different physical picture. Our results shed light on the nonlinear response of such structures using a simple and intuitive approach. 

In conclusion, we have provided a unified framework for understanding giant Kerr nonlinearities in guided-wave structures that generalizes the ENZ condition to guided wave geometries:
both possesses the largest Kerr nonlinear response where the longitudinal- and transverse- field enhancements are largest. Our insights point to new design paradigms for energy efficient and compact nanoplasmonic devices with large nonlinearities.

\noindent\textbf{Funding.} Australian Research Council Discovery Early Career Researcher Award (DE200101041).

\noindent\textbf{Disclosures.} The authors declare no conflicts of interest.

\bibliography{main_bbl}

\begin{thebibliography}{10}
\newcommand{\enquote}[1]{``#1''}

\bibitem{byer1974nonlinear}
R.~L. Byer, {\protect\JournalTitle{Annual Review of Materials Science}}
  \textbf{4}, 147 (1974).

\bibitem{chang2014quantum}
D.~E. Chang, V.~Vuleti{\'c}, and M.~D. Lukin, {\protect\JournalTitle{Nature
  Photonics}} \textbf{8}, 685 (2014).

\bibitem{mesch2016nonlinear}
M.~Mesch, B.~Metzger, M.~Hentschel, and H.~Giessen, {\protect\JournalTitle{Nano
  Letters}} \textbf{16}, 3155 (2016).

\bibitem{cotter1999nonlinear}
D.~Cotter, R.~Manning, K.~Blow, A.~Ellis, A.~Kelly, D.~Nesset, I.~Phillips,
  A.~Poustie, and D.~Rogers, {\protect\JournalTitle{Science}} \textbf{286},
  1523 (1999).

\bibitem{mukamel1999principles}
S.~Mukamel, \emph{Principles of nonlinear optical spectroscopy}, 6 (Oxford
  University Press on Demand, 1999).

\bibitem{niu2018epsilon}
X.~Niu, X.~Hu, S.~Chu, and Q.~Gong, {\protect\JournalTitle{Advanced Optical
  Materials}} \textbf{6}, 1701292 (2018).

\bibitem{boyd2019nonlinear}
R.~W. Boyd, \emph{Nonlinear optics} (Academic press, 2019).

\bibitem{caspani2016enhanced}
L.~Caspani, R.~Kaipurath, M.~Clerici, M.~Ferrera, T.~Roger, J.~Kim, N.~Kinsey,
  M.~Pietrzyk, A.~Di~Falco, V.~M. Shalaev \emph{et~al.},
  {\protect\JournalTitle{Physical Review Letters}} \textbf{116}, 233901 (2016).

\bibitem{alam2016large}
M.~Z. Alam, I.~De~Leon, and R.~W. Boyd, {\protect\JournalTitle{Science}}
  \textbf{352}, 795 (2016).

\bibitem{moitra2013realization}
P.~Moitra, Y.~Yang, Z.~Anderson, I.~I. Kravchenko, D.~P. Briggs, and
  J.~Valentine, {\protect\JournalTitle{Nature Photonics}} \textbf{7}, 791
  (2013).

\bibitem{kaipurath2016optically}
R.~Kaipurath, M.~Pietrzyk, L.~Caspani, T.~Roger, M.~Clerici, C.~Rizza,
  A.~Ciattoni, A.~Di~Falco, and D.~Faccio, {\protect\JournalTitle{Scientific
  Reports}} \textbf{6}, 27700 (2016).

\bibitem{capretti2015comparative}
A.~Capretti, Y.~Wang, N.~Engheta, and L.~Dal~Negro, {\protect\JournalTitle{ACS
  Photonics}} \textbf{2}, 1584 (2015).

\bibitem{reshef2019nonlinear}
O.~Reshef, I.~De~Leon, M.~Z. Alam, and R.~W. Boyd,
  {\protect\JournalTitle{Nature Reviews Materials}} \textbf{4}, 535 (2019).

\bibitem{campione2015theory}
S.~Campione, I.~Brener, and F.~Marquier, {\protect\JournalTitle{Physical Review
  B}} \textbf{91}, 121408 (2015).

\bibitem{vassant2012berreman}
S.~Vassant, J.-P. Hugonin, F.~Marquier, and J.-J. Greffet,
  {\protect\JournalTitle{Optics Express}} \textbf{20}, 23971 (2012).

\bibitem{campione2016experimental}
S.~Campione, I.~Kim, D.~de~Ceglia, G.~A. Keeler, and T.~S. Luk,
  {\protect\JournalTitle{Optics Express}} \textbf{24}, 18782 (2016).

\bibitem{minn2018excitation}
K.~Minn, A.~Anopchenko, J.~Yang, and H.~W.~H. Lee,
  {\protect\JournalTitle{Scientific Reports}} \textbf{8}, 2342 (2018).

\bibitem{argyropoulos2012boosting}
C.~Argyropoulos, P.-Y. Chen, G.~D’Aguanno, N.~Engheta, and A.~Alu,
  {\protect\JournalTitle{Physical Review B}} \textbf{85}, 045129 (2012).

\bibitem{li2019exceptional}
Y.~Li and C.~Argyropoulos, {\protect\JournalTitle{Physical Review B}}
  \textbf{99}, 075413 (2019).

\bibitem{li2018epsilon}
Y.~Li and C.~Argyropoulos, {\protect\JournalTitle{Active Photonic Platforms X}}
  \textbf{10721}, 1072106 (2018).

\bibitem{li2017general}
G.~Li, C.~M. de~Sterke, and S.~Palomba, {\protect\JournalTitle{Optics Letters}}
  \textbf{42}, 1329 (2017).

\bibitem{li2020establishing}
G.~H.~Y. Li, A.~Tuniz, and C.~M. de~Sterke, {\protect\JournalTitle{Optics
  Letters}}  (in press).

\bibitem{kinsey2019nonlinear}
N.~Kinsey and J.~Khurgin, {\protect\JournalTitle{Optical Materials Express}}
  \textbf{9}, 2793 (2019).

\bibitem{tuniz2020modular}
A.~Tuniz, O.~Bickerton, F.~J. Diaz, T.~K{\"a}sebier, E.-B. Kley, S.~Kroker,
  S.~Palomba, and C.~M. de~Sterke, {\protect\JournalTitle{Nature
  Communications}} \textbf{11}, 1 (2020).

\bibitem{maier2007plasmonics}
S.~A. Maier, \emph{Plasmonics: fundamentals and applications} (Springer Science
  \& Business Media, 2007).

\bibitem{afshar2013understanding}
V.~S. Afshar, T.~Monro, and C.~M. de~Sterke, {\protect\JournalTitle{Optics
  Express}} \textbf{21}, 18558 (2013).

\bibitem{Monat2010review}
C.~Monat, C.~M. de~Sterke, and B.~J. Eggleton, {\protect\JournalTitle{J. Opt}}
  \textbf{12}, 104003 (2010).

\bibitem{miyata2012field}
M.~Miyata and J.~Takahara, {\protect\JournalTitle{Journal of Applied Physics}}
  \textbf{111}, 053102 (2012).

\bibitem{schmidt2008long}
M.~Schmidt and P.~S.~J. Russell, {\protect\JournalTitle{Optics Express}}
  \textbf{16}, 13617 (2008).

\bibitem{novotny1994light}
L.~Novotny and C.~Hafner, {\protect\JournalTitle{Physical Review E}}
  \textbf{50}, 4094 (1994).

\bibitem{tuniz2020pulse}
A.~Tuniz, S.~Palomba, and C.~M. de~Sterke, {\protect\JournalTitle{Applied
  Physics Letters}} \textbf{117}, 071105 (2020).

\bibitem{boyd1989applications}
G.~Boyd, {\protect\JournalTitle{J. Opt. Soc. Am. B}} \textbf{6}, 685 (1989).

\bibitem{reshef2017beyond}
O.~Reshef, E.~Giese, M.~Z. Alam, I.~De~Leon, J.~Upham, and R.~W. Boyd,
  {\protect\JournalTitle{Optics Letters}} \textbf{42}, 3225 (2017).

\bibitem{guo2016large}
P.~Guo, R.~D. Schaller, L.~E. Ocola, B.~T. Diroll, J.~B. Ketterson, and R.~P.
  Chang, {\protect\JournalTitle{Nature Communications}} \textbf{7}, 1 (2016).

\end{thebibliography}

\bibliographyfullrefs{sample}


\end{document}